\documentclass[conference]{IEEEtran}
\IEEEoverridecommandlockouts
\usepackage[numbers,sort&compress]{natbib}
\usepackage{amsmath,amssymb,amsfonts}
\usepackage{algorithmic}
\usepackage{graphicx}
\usepackage{textcomp}
\usepackage{xcolor}

\usepackage{multirow}
\usepackage{booktabs}

\def\BibTeX{{\rm B\kern-.05em{\sc i\kern-.025em b}\kern-.08em
    T\kern-.1667em\lower.7ex\hbox{E}\kern-.125emX}}
\begin{document}

\title{OpTI-Mouse: Optimization for Targeted Temporal Interference Stimulation in the Mouse Brain\\
\thanks{This work was supported by the National Key R\&D Program of China (2025YFC3410000), National Natural Science Foundation of China (62472206, 3254100307), Shenzhen Science and Technology Innovation Committee (RCYX20231211090405003, JCYJ20220818100213029), Guangdong Provincial Key Laboratory of Advanced Biomaterials (2022B1212010003), and the open research fund of the Guangdong Provincial Key Laboratory of Mathematical and Neural Dynamical Systems, the Center for Computational Science and Engineering at Southern University of Science and Technology. We would like to thank ZMT Zurich MedTech AG for providing Sim4Life software.
}
\thanks{\footnotesize Accepted for publication in \emph{Proc. IEEE EMBC 2026}.}
}

\author{
   \IEEEauthorblockN{
       Jingsheng Tang\textsuperscript{1}\IEEEauthorrefmark{2},
       Zhengkang Zhou\textsuperscript{1}\IEEEauthorrefmark{2},
       Yingyue Xin\textsuperscript{1},
       Zihan Ning\textsuperscript{1},
       Pengfei Wei\textsuperscript{2},
       Mo Wang\textsuperscript{1}\IEEEauthorrefmark{1},
       Quanying Liu\textsuperscript{1}\IEEEauthorrefmark{1}
   }
   \IEEEauthorblockA{   \textsuperscript{1}Department of Biomedical Engineering, Southern University of Science and Technology, Shenzhen, China\\
   \textsuperscript{2}School of Biological Science and Medical Engineering, State Key Laboratory of Digital Medicine,\\ Southeast University, Nanjing, China\\
   \IEEEauthorrefmark{2}These authors contributed equally\\
   \IEEEauthorrefmark{1}Corresponding authors: Quanying Liu (liuqy@sustech.edu.cn) and Mo Wang (12250099@mail.sustech.edu.cn)
   }
}


\maketitle

\begin{abstract}
Temporal Interference (TI) stimulation enables deep brain targeting, yet precise optimization tools for mouse models remain limited. We developed a computational optimization tool integrating mouse head modeling with the optimization algorithm to optimize stimulation strategies for predefined target regions. By balancing target intensity and spatial focality, the optimized strategy significantly outperformed empirical baselines. For the CA3-CA1 target, it achieved a 7-fold intensity increase (10.29 vs. 2.89 V/m) under iso-focality conditions. Conversely, for the Dentate Gyrus, it improved spatial confinement ($r_{0.5}$ reduced from 3.99 to 3.54 mm) while maintaining comparable intensity. Cross-model validation on a standardized Sim4Life phantom further confirmed the framework's robustness. This approach offers a powerful tool for enhancing the precision and reproducibility of preclinical TI stimulation studies. 
\end{abstract}


\section{Introduction}
Temporal Interference (TI) stimulation is an emerging non-invasive neuromodulation technique designed to target deep brain structures while sparing superficial regions~\cite{grossman2017noninvasive}. Owing to its high targeting precision, TI stimulation has shown immense promise for the treatment of human neurological disorders~\cite{missey2025non,yang2025transcranial}. Nevertheless, the precise physiological effects and underlying neural mechanisms of TI in the human brain remain largely elusive. Therefore, murine models have been increasingly adopted to dissect therapeutic mechanisms and investigate brain-behavior causality~\cite{qi2024temporally,liu2025intraoperative,wang2025transcranial}. Despite this wide adoption, current murine protocols rely heavily on empirical settings~\cite{grossman2017noninvasive,missey2021orientation,acerbo2024improved}, which do not guarantee an optimal balance between target intensity and spatial focality. This necessitates the development of advanced computational frameworks to systematically optimize stimulation parameters for precise and reproducible modulation~\cite{wang2026personalized}. 


TI stimulation works by superimposing two high-frequency carriers with a small frequency difference to create a low-frequency envelope at the target. Optimizing this process to maximize target modulation amplitude is a constrained non-convex problem~\cite{huang2020optimization}. Various computational approaches have been employed to address this optimization problem, ranging from hybrid linear programming and sequential quadratic programming~\cite{huang2020optimization,huang2019can} to unsupervised neural networks~\cite{bahn2022computational} and genetic algorithms~\cite{stoupis2022non}. Notably, comprehensive automated pipelines such as SimNIBS~\cite{saturnino2019simnibs} and ROAST~\cite{huang2018roast} have been developed to streamline the workflow from forward modeling to inverse optimization. However, these tools are tailored primarily for human head models, leaving a significant gap in dedicated frameworks that seamlessly integrate forward modeling with inverse optimization for mouse research. 

\begin{figure*}[t]
    \centering
    \includegraphics[width=0.95\linewidth]{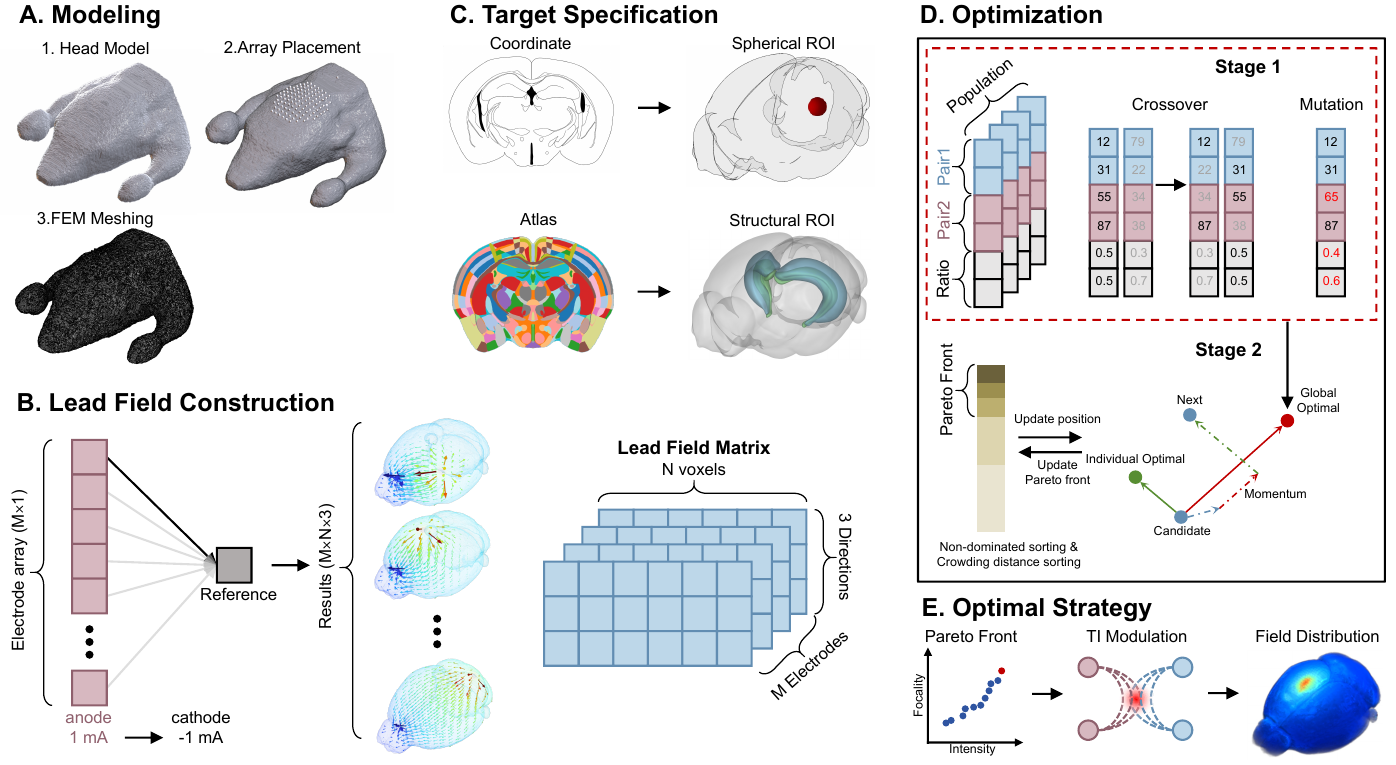}
    \caption{\textbf{Overview of OpTI-Mouse: a computational framework for targeted TI stimulation in the mouse brain.} \textbf{(A)} Construction of the forward model, including high-density electrode array placement and finite element meshing. \textbf{(B)} Calculation of the lead field matrix to map electrode currents to intracerebral electric fields. \textbf{(C)} Definition of target regions of interest (ROI) using either spherical ROI with stereotaxic coordinates (up) or structural ROI in anatomical atlas (bottom). \textbf{(D)} The multi-objective optimization process (MOVEA~\cite{wang2023multi}), which synergizes a Genetic Algorithm (Stage 1) for global search with Particle Swarm Optimization (Stage 2) to iteratively evolve the Pareto front by identifying non-dominated solutions. \textbf{(E)} Visualization of the optimal strategy, illustrating the trade-off between target intensity and focality alongside the resulting focused TI field distribution.}
    \label{fig:framework}
\end{figure*}

In this study, we presented a computational optimization tool for TI stimulation, called OpTI-Mouse. It combines head conductivity volume modeling of the mouse brain (i.e., forward modeling) and an optimization strategy (i.e., inverse optimization) to search for optimal stimulation strategies for predefined targeted regions in the mouse brain (Fig.~\ref{fig:framework}). Subsequently, to test the efficacy of OpTI-Mouse, we quantitatively compared optimized TI stimulation from OpTI-Mouse against the stimulation strategies from conventional empirical methods. Finally, we assessed the consistency of the results across different anatomical models to validate the robustness and generalizability of the optimization strategy. Our main contributions are summarized as follows.

\begin{enumerate}
    \item We developed OpTI-Mouse, the first computational tool adaptable to various mouse head models, which integrates inverse optimization for targeted TI stimulation. 
    \item The superior efficacy of OpTI-Mouse is demonstrated by comparing with conventional empirical strategies.
    \item The generalizability and robustness of OpTI-Mouse are quantitatively evaluated on different mouse models. 
\end{enumerate}

\section{Methods}



\subsection{Forward modeling}
To establish a realistic mouse head model, as shown in Fig.~\ref{fig:framework}A, the TMBTA brain model~\cite{barriere2021brain} was coregistered with skull~\cite{chan2007development} and scalp~\cite{dogdas2007digimouse} structures, including a cerebrospinal fluid (CSF) layer. A custom Python script automated the precise placement of electrodes (radius: 0.25 mm; center-to-center distance: 1 mm). Subsequently, the complete head model was meshed from NIfTI data using Iso2mesh~\cite{tran2020improving}. 

The electric field distribution within the head model was calculated using the Finite Element Method (FEM) implemented in COMSOL Multiphysics 6.3 (COMSOL, Inc., Burlington, MA). Specifically, we utilized the Electric Currents interface with a stationary solver. Upon importing the mesh, isotropic electrical conductivities were assigned to the tissues as follows: gray matter (0.275 S/m), cerebrospinal fluid (CSF; 1.654 S/m),  skull (0.01 S/m), and scalp (0.465 S/m).

Under quasi-static conditions, and assuming no internal current sources within the brain, the electric potential distribution within the volume conductor is governed by the Laplace equation:
\begin{equation}
    \nabla \cdot \mathbf{J} = \nabla \cdot (\sigma \mathbf{E}) = -\nabla \cdot (\sigma \nabla V) = 0
\end{equation}
where $\sigma$ represents the electrical conductivity (S/m), $V$ denotes the electric potential (V), $\mathbf{E}$ is the electric field vector (V/m), and $\mathbf{J}$ represents the current density vector (A/m$^2$). To accurately model constant current stimulation, we apply a global Neumann boundary condition at the active electrode surfaces:
\begin{equation}
    \int_{\partial \Omega} \mathbf{J} \cdot \mathbf{n} \, dS = I_0
\end{equation}
where $\partial \Omega$ denotes the electrode-tissue interface, $\mathbf{n}$ is the outward unit normal vector, and $I_0$ represents the total injected current amplitude (e.g., 1 mA).

To construct the lead field matrix ($L$), a unit current was sequentially applied to each electrode in the array (acting as the anode) relative to the reference electrode (acting as the cathode). The lead field matrix encodes the electric field intensity generated by each individual electrode at every spatial location within the brain. This matrix enables the rapid calculation of the electric field distribution for any pair of electrodes (Fig.~\ref{fig:framework}B):
\begin{equation}
    E = Lx \cdot (\vec{n})
\end{equation}
where $x$ denotes the stimulation configuration matrix (activated channels), and $\vec{n}$ is the directional vector corresponding to the target orientation. 

For TI stimulation, the modulation envelope was calculated using the formula proposed by Huang et al.~\cite{huang2020optimization}:
\begin{equation}
|\vec{E}(\vec{r})| = \left| \left|\left(\vec{E}_1(\vec{r}) + \vec{E}_2(\vec{r})\right) \cdot \vec{n}\right| - \left|\left(\vec{E}_1(\vec{r}) - \vec{E}_2(\vec{r})\right) \cdot \vec{n}\right| \right|
\end{equation}
where $\vec{E}_1(\vec{r})$ and $\vec{E}_2(\vec{r})$ represent the electric field vectors at position $\vec{r}$ generated by the first and second electrode pairs, respectively, and $\vec{n}$ denotes the unit vector along the target direction.

Since the optimal orientation for neuronal activation is often uncertain, we computed the modulation depth along the direction that yields the maximal magnitude:
\begin{equation}
|\vec{E}|_{\text{max}} = 2 \max_{\alpha} \min\left( \left\|\vec{E}_{1}\right\| |\cos \alpha|, \left\|\vec{E}_{2}\right\| |\cos (\alpha-\phi)| \right)
\end{equation}
where $\left\|\vec{E}_{1}\right\|$ and $\left\|\vec{E}_{2}\right\|$ denote the magnitudes of the electric field vectors generated by the two electrode pairs, $\phi$ represents the spatial angle between these two vectors, and $\alpha$ is the orientation angle of the projection axis relative to $\vec{E}_{1}$. 

\subsection{TI Optimization}
We formulated an inverse optimization problem to determine the optimal electrode configuration and TI stimulation parameters for a given target location. Due to the inherent trade-off between maximizing target intensity and focality, we leveraged the Multi-objective optimization via evolutionary algorithm (MOVEA)~\cite{wang2023multi} to optimize these competing objectives subjective to three safety constraints:
\begin{align}
g_{1}(s) &= \sum_{n} |s_{n}| + \left|\sum_{n} s_{n}\right| \leq 2I_{\text{tot}},\label{eq:safety_total}\\
g_{2}(s) &= |s_{n}| \leq I_{\text{ind}},\label{eq:safety_ind_n}\\
g_{3}(s) &= \left|\sum_{n} s_{n}\right| \leq I_{\text{ind}}\label{eq:safety_ind_sum},
\end{align}
where $s_{n}$ denotes the current amplitude applied to the $n$-th electrode, $I_{\text{tot}}$ represents the maximum allowable total injected current, and $I_{\text{ind}}$ is the safety threshold for the current intensity at any individual electrode. 

The optimization problem was defined with two competing objectives: maximizing target field intensity and minimizing the half-max radius ($r_{0.5}$) as a measure of focality:
\begin{align}
    \max f(s) &= \sum_{i \in \mathcal{T}} |\vec{E}|_{\text{max},i},\\
    \min f(s) &= r_{0.5}
\end{align}
where $\mathcal{T}$ denotes the set of indices corresponding to the target region of interest (ROI). $|\vec{E}|_{\text{max},i}$ indicates the maximum electric field intensity at the $i$-th index. $r_{0.5}$ represents the radius of the spherical region centered at the target where the electric field intensity exceeds 50\% of the whole-brain maximum.

Finally, we employed MOVEA to optimize the stimulation strategy. This hybrid framework synergizes Genetic Algorithms (GA) with Particle Swarm Optimization (PSO). Specifically, as illustrated in Fig.~\ref{fig:framework}D, a single-objective GA is first executed to identify the solution yielding maximum target intensity. This solution subsequently serves as prior knowledge to initialize the Multi-Objective Particle Swarm Optimization (MOPSO), which explores the Pareto front to balance the competing objectives.

For a new target region, users first specify the ROI either using stereotaxic coordinates or atlas-based anatomical segmentation (Fig.~\ref{fig:framework}C). The dense electrode array is retained as a candidate search space, and the optimizer automatically adapts electrode-pair combinations and current ratios to the specified ROI under the same current constraints. Thus, regional adaptation is achieved through ROI redefinition and inverse optimization, while the final montage should still be checked according to practical electrode accessibility and placement constraints.

\begin{figure*}[t]
    \centering
    \includegraphics[width=0.9\textwidth]{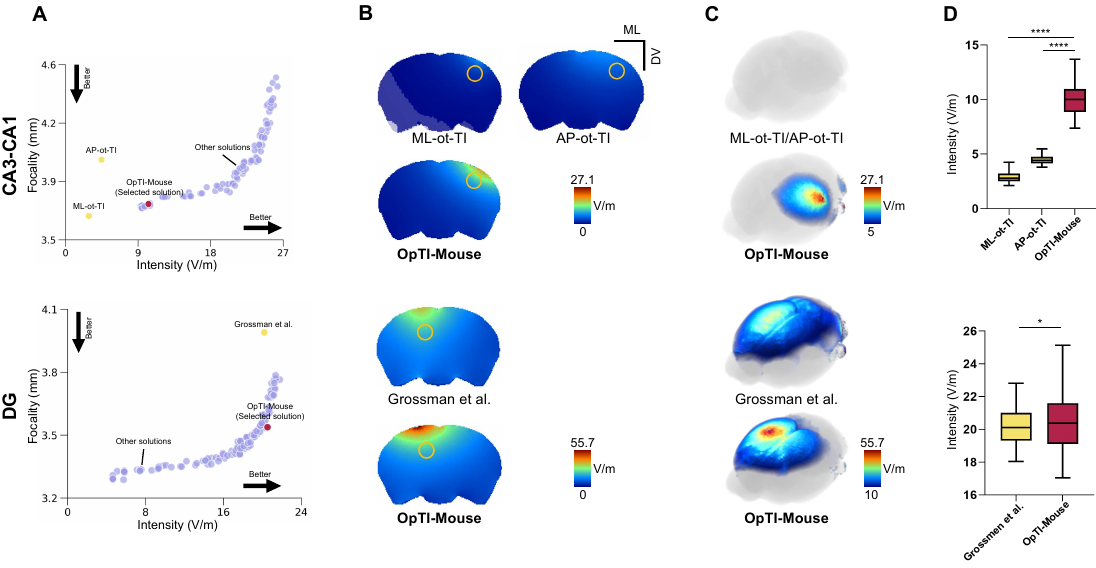}
    \caption{\textbf{Quantitative and qualitative comparison of optimized stimulation strategies against empirical baselines.} (A) Pareto Fronts of TI optimization for targeting area in CA3-CA1 (top) or in DG (bottom). Light violet dots indicate all the optimal solutions in Pareto Fronts. These Pareto fronts show a trade-off between target intensity and spatial focality. The highlighted yellow dots and the red dot denote the empirical baselines and the corresponding optimized strategy via OpTI-Mouse, respectively (i.e., ML-ot-TI, AP-ot-TI, and Grossman et al.\cite{grossman2017noninvasive}).  (B) Coronal cross-sections of electric field distributions passing through the target centroids (indicated by circles) for the CA3-CA1 (AP: -1.94 mm) and DG (AP: -2.00 mm) regions. The optimized strategy demonstrates enhanced target coverage compared to baseline montages. (C) 3D visualization of activation regions exceeding specific thresholds (5 V/m for CA3-CA1; 10 V/m for DG), highlighting the superior spatial confinement of the proposed method. (D) Statistical analysis of target intensity. The optimized strategy yields significantly higher field strengths compared to baselines on targets. (*$p < 0.05$, ****$p < 0.0001$)}
    \label{fig:result1}
\end{figure*}

\section{Experiments}
\subsection{Experimental Setup}
All forward modeling and optimization experiments were conducted on a computing platform equipped with an Intel Core i7-10700K CPU, 128 GB RAM, and an NVIDIA GeForce RTX 3080 GPU. Finite element simulations were carried out using both COMSOL and Sim4Life (ZMT, Zurich, Switzerland)~\cite{sim4life}, while the optimization algorithms were executed in a custom Python 3.8 environment. To ensure a rigorous and fair comparison, identical computational conditions and target definitions were strictly maintained across all simulations. 

\subsection{Empirical TI Stimulation Setups}
Two challenging deep-brain targets of the hippocampus were selected to evaluate the optimization framework: the CA3-CA1 region (AP = -1.94 mm, ML = -2.80 mm, DV = 1.57 mm) and the Dentate Gyrus (DG) (AP = -2.00 mm, ML = 1.00 mm, DV = 2.00 mm). To benchmark our OpTI-Mouse, three baseline TI stimulation strategies were constructed based on empirically designed TI configurations reported in prior work~\cite{grossman2017noninvasive,missey2021orientation}. Baselines 1 and 2 (Target: CA3-CA1) adopted the strategies described by Missey et al.~\cite{missey2021orientation}, employing fixed electrode arrangements aligned along anatomical axes. Specifically, Baseline 1 aligned electrodes along the mediolateral axis (AP = -1.94 mm; ML = +0.5, -0.5, -3.9, -4.3 mm), while Baseline 2 distributed electrodes along the anterior--posterior axis (ML = -2.04 mm; AP = +2.2, +1.1, -4.4, -5.5 mm). Baseline 3 (Target: DG) reproduced the paradigm introduced by Grossman et al.~\cite{grossman2017noninvasive}, positioning cranial electrodes at AP = -2 mm (ML = -0.25 and 2.75 mm) paired with a ventral trunk patch. These empirical configurations were evaluated via forward simulations with the total injected current set to 1 mA to serve as reference standards for quantitative comparison.

\subsection{Optimization}

All optimized configurations in this study targeted a spherical ROI with a radius of 0.5 mm, subject to a fixed total current limit of 1 mA. The optimization followed a two-stage hybrid strategy (Fig.~\ref{fig:framework}D): (1) GA was first deployed to maximize target intensity and provide high-quality candidate solutions for subsequent multi-objective refinement; (2) MOPSO was then used to iteratively evolve the Pareto front, balancing intensity and focality. In this implementation, GA was run with a population size of 50 for 50 generations, and MOPSO was run with 100 particles for 500 iterations. These hyperparameters were empirically selected as a practical trade-off among optimization quality, Pareto-front sampling density, and computational complexity. To account for algorithmic stochasticity, each experiment was repeated five times. Finally, a representative optimal strategy was selected from the resulting Pareto front for quantitative comparison with the baseline configurations (Fig.~\ref{fig:framework}E).

\subsection{Cross-Model Validation}

Cross-model simulations were performed to validate the mouse head model constructed in this study by comparison with a reference implementation in Sim4Life.

The reference framework was implemented using the C57BL/6N\_F\_E19 anatomical model (ViZoo library)~\cite{murbach2018weiterfuehrende}. Simulations assumed the ohmic quasi-static approximation with constant-current boundary conditions (1 mA) at paired source--sink electrodes. Tissue properties followed the IT'IS LF 4.1 database. The simulation domain was cropped to 20~mm $\times$ 25~mm $\times$ 40~mm and meshed with a geometric resolution of 0.05~mm. TI modulation amplitudes were solved using the \textit{MaxModulation} module in Sim4Life. Finally, spatial consistency was assessed by comparing the electric field and TI modulation envelope distributions under identical stimulation strategies. 

\section{Results and Discussion}

\subsection{Pareto Optimization Analysis}
The optimization results for both CA3-CA1 and DG targets are visualized as Pareto fronts in Fig.~\ref{fig:result1}A, illustrating the set of non-dominated solutions yielded by our framework. These pareto fronts reveal a characteristic trade-off between the two conflicting objectives: maximizing target intensity imposes a cost on spatial focality. The optimization yielded a broad spectrum of non-dominated solutions for both targets: CA3-CA1 (Intensity: 9.29--26.28 V/m; Focality: 3.70--4.52 mm) and DG (Intensity: 4.64--21.78 V/m; Focality: 3.29--3.79 mm). 

To compare optimization efficacy with empirical standards, the CA3-CA1 strategy was selected from the Pareto front region where focality values converged with those of the baselines. In contrast, the DG strategy was chosen from the iso-intensity region, matching the field strength of the baseline configuration. 

\subsection{Comparison with Empirical Baselines}\label{result-comparison}
Visualization of the electric field distributions reveals differences in targeting performance. For the CA3-CA1 target, under identical current constraints, the empirical electrode montages generated overall envelope modulation amplitudes that were substantially lower than those of the optimized strategy, resulting in notably weaker target intensity (Fig.~\ref{fig:result1}B, C, top). Conversely, for the DG target, while the empirical baseline achieved target intensities comparable to the optimized solution (Fig.~\ref{fig:result1}B, bottom), it demonstrated poor spatial convergence. As shown in Fig.~\ref{fig:result1}C, this resulted in a larger spatial extent exceeding the activation threshold (e.g., 10 V/m) compared with the optimized approach.

Quantitative analysis confirmed these performance gains. For the CA3-CA1 target, the optimized strategy achieved a robust increase in mean target intensity to 10.29 V/m while maintaining focality comparable to the ML-ot-TI baseline (3.65 mm) (Table~\ref{tab1}). This represents a 7-fold and 4.5-fold enhancement over the ML-ot-TI (2.89 V/m) and AP-ot-TI (4.47 V/m) configurations, respectively. Statistical analysis confirmed that the optimized intensity was significantly higher than both empirical groups (Kruskal-Wallis test: $H=1383, p < 0.0001$; Dunn's post-hoc: $p < 0.0001$ for both comparisons; Fig.~\ref{fig:result1}D, top).

For the DG target, OpTI-Mouse balanced the competing objectives of intensity and focality. The selected strategy achieved an intensity level comparable to the baseline (20.52 V/m vs. 20.19 V/m) (Table~\ref{tab1}). Although statistical analysis detected a significant difference due to the low variance of the datasets (Mann-Whitney U test: $p=0.0393$, Fig.~\ref{fig:result1}D, bottom), the absolute magnitude of this increase was marginal ($<1.5\%$). Under these matched field-strength conditions, the optimized configuration reduced $r_{0.5}$ to 3.54 mm compared with 3.99 mm for the baseline (Table~\ref{tab1}), indicating improved spatial selectivity.

\begin{table}[!t]
\centering
\caption{Quantitative comparison of stimulation strategies. Intensity: target electrical field; Focality: half-max radius ($r_{0.5}$). }
\label{tab1}
\resizebox{0.48\textwidth}{!}{%
\begin{tabular}{c c c c}
\toprule
\textbf{Target} & \textbf{Strategy} & \textbf{Intensity} (V/m) $\uparrow$ & \textbf{Focality} (mm) $\downarrow$ \\
\midrule

\multirow{3}{*}{CA3-CA1}
  & ML-ot-TI & 2.89 & \textbf{3.65} \\
  & AP-ot-TI & 4.47 & 4.00 \\
  & \textbf{OpTI-Mouse (ours)} & \textbf{10.29} & 3.73 \\
\midrule

\multirow{2}{*}{DG}
  & Grossman et al. & 20.19 & 3.99 \\
  & \textbf{OpTI-Mouse (ours)} & \textbf{20.52} & \textbf{3.54} \\

\bottomrule
\end{tabular}}
\end{table}

\subsection{Generalizability Assessment}

To assess the generalizability of the optimization strategies across different mouse head models, we conducted spatial and quantitative analyses of the stimulation outcomes in the CA3–CA1 and DG target regions using the Sim4Life reference model. Representative solutions derived from the OpTI-Mouse pipeline and the empirical strategies were directly applied to the reference model to evaluate performance consistency. 

\begin{figure}
    \centering
    \includegraphics[width=\columnwidth]{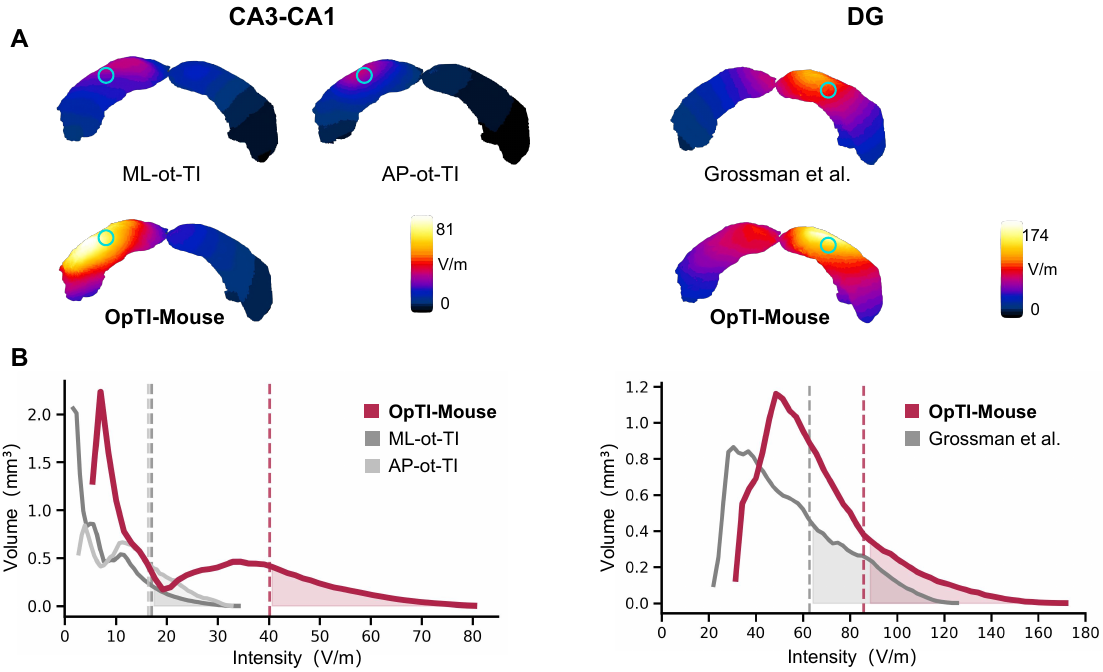}
    \caption{\textbf{\textbf{Spatial and quantitative comparison of stimulation strategies in the reference mode}} (A) Surface projections of TI modulation magnitude mapped onto the hippocampus. Circles indicate the target centroids. For the CA3–CA1 target (left column), the upper and lower panels show the ML-ot-TI and AP-ot-TI empirical baselines, respectively, while the right panel shows the optimized OpTI-Mouse strategy. For the DG target (right column), the left panel shows the empirical strategy reported by Grossman et al.~\cite{grossman2017noninvasive}, and the right panel shows the optimized strategy. All projections are generated from simulations performed in the reference model under identical current constraints.
(B) Volume–intensity curves characterizing whole-hippocampus exposure for stimulation configurations optimized for different targets. The curves depict the volumetric distribution of modulation intensity across the hippocampus, quantifying the tissue volume associated with each intensity level. Vertical dashed lines indicate half of the 95th percentile of the corresponding intensity distributions. }  
    \label{fig:result2}
\end{figure}

As illustrated in Fig.~\ref{fig:result2}, the optimized strategy delivers significantly higher TI envelope amplitudes in both target regions, achieving enhanced field coverage over the ROIs. In contrast, under identical simulation conditions, the empirical strategies exhibit diminished field intensity and substantially lower modulation magnitudes. 

The corresponding quantitative results are summarized in Table~\ref{tab2}.
For the CA3--CA1 target region, the optimized strategy achieves a target-region intensity of 35.88 V/m in the reference model, substantially exceeding ML-ot-TI (9.25 V/m) and AP-ot-TI (11.97 V/m), while maintaining a spatial focality comparable to the empirical strategies. For the DG target region, the optimized strategy yields a higher target-region intensity (167.18 V/m) than the empirical strategy (121.68 V/m), accompanied by an improvement in spatial focality ($r_{0.5}$) from 3.07 mm to 2.29 mm. Considering the volume-intensity results of whole-hippocampus (Fig.~\ref{fig:result2}B), although optimization yields an increase in off-target intensity, the distinct advantage in focality permits the use of reduced stimulation currents to maintain the activation volume within a controllable range. 

Comparing the findings with Section~\ref{result-comparison}, we observe that although the qualitative conclusions regarding strategy efficacy remain consistent, the quantitative magnitudes differ. This variation likely stems from: (1) registration differences, as the reference model was not aligned to a stereotactic atlas space, likely introducing electrode positioning errors; and (2) resolution disparity. Specifically, the finer segmentation resolution of our model (0.066 mm $\times$ 0.066 mm $\times$ 0.066 mm) versus the coarser reference model (0.0825 mm $\times$ 0.077 mm $\times$ 0.075 mm) resulted in varying mesh densities during discretization, thereby leading to distinct finite element solutions. 

Taken together, the relative performance trends among different stimulation strategies remained largely consistent within the reference model. These consistent findings underscore the robust generalizability of the strategies generated by OpTI-Mouse, while also highlighting the tool's potential for cross-model application.

\begin{table}[htbp]
\centering 
\caption{Quantitative comparison of stimulation strategies using the reference model. Intensity: target electrical field; Focality: half-max radius ($r_{0.5}$).} 
\label{tab2} 
\resizebox{0.48\textwidth}{!}{%
\begin{tabular}{c c c c} 
\toprule
\textbf{Target} & \textbf{Strategy} & \textbf{Intensity} (V/m) $\uparrow$ & \textbf{Focality} (mm) $\downarrow$ \\ 
\midrule 

\multirow{3}{*}{CA3-CA1} & ML-ot-TI & 9.25 &6.50 \\ & AP-ot-TI & 11.97 & 6.62 \\ & \textbf{OpTI-Mouse (ours)} & \textbf{35.88} & \textbf{6.39} \\ \midrule 

\multirow{2}{*}{DG} & Grossman et al. & 121.68 &3.07 \\ & \textbf{OpTI-Mouse (ours)} & \textbf{167.18} & \textbf{2.29} \\ 

\bottomrule \end{tabular}} \end{table}

\section{Conclusion}
In summary, we developed OpTI-Mouse, a computational framework for optimizing TI stimulation in the mouse brain. Quantitative comparisons demonstrated that stimulation strategies optimized via OpTI-Mouse markedly outperformed empirical baselines by effectively resolving the trade-off between intensity and focality. Validated by cross-model simulations, OpTI-Mouse provides a robust and automated planning tool to enhance the precision and reproducibility of preclinical neuromodulation research.

\bibliographystyle{IEEEtran}
\bibliography{ref.bib}

\end{document}